\documentclass[hidelinks,fleqn,10pt]{olplainarticle}
\usepackage{subcaption}
\usepackage{bbm}
\usepackage{xurl}
 \usepackage{orcidlink}

\newtheorem{proposition}{Proposition}
\newcommand{\R}{\mathbb{R}}
\newcommand{\E}{\mathbb{E}}
\newcommand{\Tr}{\text{Tr}}
\DeclareMathOperator*{\argmin}{arg\,min}
\def\qed{\hfill$\blacksquare$\\} \newenvironment{proof}{\noindent {\bf
Proof.}}{\qed}
\newtheorem{innercustomthm}{Proposition}
\newenvironment{myprop}[1]
  {\renewcommand\theinnercustomthm{#1}\innercustomthm}
  {\endinnercustomthm}

\title{Hierarchical Bayesian Models to Mitigate Systematic Disparities in Prediction with Proxy Outcomes}

\author[1]{Jonas Mikhaeil\, \orcidlink{0000-0001-6745-7505}
\footnote{Address for correspondence: Department of Statistics, 
Columbia University,
New York 10027.
Email: j.mikhaeil@columbia.edu}}
\author[1,2]{Andrew Gelman}
\author[1]{Philip Greengard}

\affil[1]{Department of Statistics, Columbia University, New York}
\affil[2]{Department of Political Science, Columbia University, New York}


\begin{abstract}

Label bias occurs when the outcome of interest is not directly observable and instead, modeling is performed with proxy labels. When the difference between the true outcome and the proxy label is correlated with predictors, this can yield systematic disparities in predictions for different groups of interest. 
We propose Bayesian hierarchical measurement models to address these issues. 
When strong prior information about the measurement process is available, our approach improves accuracy and helps with algorithmic fairness. If prior knowledge is limited, our approach allows assessment of the sensitivity of predictions to the unknown specifications of the measurement process. This can help practitioners gauge if enough substantive information is available to guarantee the desired accuracy and avoid disparate predictions when using proxy outcomes.
We demonstrate our approach through practical examples.
\end{abstract}

\date{20 Feb 2024}

\begin{document}

\flushbottom
\maketitle
\thispagestyle{empty}
\section{Introduction}

In the social sciences, measurement is often indirect, and researchers use proxy outcomes \citep{Adcock_Collier_2001, Knox_Lucas_Cho_2022}.
Even seemingly objective outcomes such as suicide rates can be systematically distorted \citep{Douglas_1967}. Sociological accounts of the processes with which data are collected highlight the unavoidable imperfections of data more broadly \citep{STARR}.
The use of imperfect proxies for the outcome can reduce the accuracy of predictions that are relevant to downstream decisions, possibly underserving specific subgroups of the population \citep{obermeyer,Fogliato_G’Sell_Chouldechova_2020,Mullainathan_Obermeyer_2021}. We propose to mitigate these problems by modeling the relationship between proxy and true outcomes with Bayesian measurement models.

Consider the example of building a statistical model to predict diabetes risk using demographic and health information from survey data. 
The goal of building such a model is to be able to cheaply identify patients who are at risk of diabetes and who should undergo more costly and time-consuming testing. The model should be accurate and calibrated.  If the model underpredicts the risk for certain groups of people, then decisions based on it can lead to these groups being underserved.

One challenge in this example is that we are only given the diagnosis, not true underlying disease status. There are several potential sources of error (usually referred to as \textit{label bias}) that this proxy outcome may introduce into a model. If the measurement error---the difference between the proxy outcome (survey response) and the true outcome (being diabetic)---is correlated with a predictor, then prediction errors can be correlated with that predictor. We demonstrate with a simple example in a linear regression setting in Section \ref{sec:labbias-lin-reg} and return to the example of diabetes risk in Section \ref{sec:diabetes}.

There are various ways of dealing with label bias in specific contexts \citep{Jiang_Nachum_2020,Wang_Liu_Levy_2021,Knox_Lucas_Cho_2022}.  Label bias often degrades prediction accuracy and, when the measurement errors are correlated with the covariates, leads to systematic errors in prediction. In the context of predicting risk, these systematic disparities in prediction are referred to as miscalibration \citep{Rothblum_Yona_2023} and have been shown to negatively impact the utility of downstream decisions \citep{Van_Calster_Vickers_2015,Parastouei_Sepandi_Eskandari_2021}.
Label bias is especially problematic when measurement errors are correlated with membership to a protected group, which is often the case in social science applications 
\citep{Biderman_Reiss_1967,Fang_Wang_Coresh_Selvin_2022,Zanger-Tishler_2023} or the healthcare sector \citep{Eneanya_Yang_Reese_2019,Cerdeña_Plaisime_Tsai_2020,Diao_Wu_Taylor_Tucker_Powe_Kohane_Manrai_2021,Basu_2023}. In this situation, decisions based on these predictions can lead to some communities being under-served on average, thus violating certain conceptions of algorithmic fairness \citep{Dwork_Hardt_Pitassi_Reingold_Zemel_2011,Hardt_Price_Price_Srebro_2016,JMLR:v24:22-1511}.

\citet{Zanger-Tishler_2023} show that in the presence of label bias, the addition of features
may deteriorate prediction accuracy on the true labels of interest. In particular, 
if a feature's correlation with the true outcome and proxy outcome, conditional on the other covariates, have different signs, then including that feature in a regression will deteriorate predictive accuracy. 
This can occur when a feature is only weakly related with the true outcome but both this feature and the outcome are causally constitutive of the remaining features.
\cite{Zanger-Tishler_2023} demonstrate this situation with the relationship between criminal behavior (the outcome of interest), arrests (the proxy outcome), and the level of policing in a neighborhood; we continue studying this example in Section \ref{Sec:leakageModel}.

In the present paper, we demonstrate that, in the setting where dropping a predictor would increase prediction accuracy, we can increase prediction accuracy even further using a measurement model and that, with sufficient knowledge about the data-generating process, measurement models can mitigate systematic disparities in prediction. 
Our work highlights the benefits of making explicit assumptions about measurement errors, even in purely predictive settings. Measurement models are a way to make these assumptions transparent and allow users to critically question if enough domain knowledge is at hand to make the proxies valid and to ensure that downstream decisions based on them do not underserve specific groups of interest. While measurement models, in principle,  allow researchers to adjust predictions to mitigate disparities and achieve decisions that improve outcomes for particular groups, the inclusion of membership information to protected groups may be problematic in itself \citep{Goel_Perelman_Shroff_Sklansky_2017} and violate the legal doctrine of ``no disparate treatment.'' We do not address this tension here; in any application with label bias of this sort, both societal and legal considerations will be crucial.  

Building measurement models tailored to specific applications has been made easier by recent advances in probabilistic programming languages such as Stan \citep{stan}, where reasonably general Bayesian models can be set up in simple, user-friendly language, allowing researchers  to represent prior knowledge, including uncertainty, about the measurement process and any discrepancy between the proxy and the true outcome in a statistical model.\footnote{All models and code to reproduce our results are available under \\
\url{https://github.com/JonasMikhaeil/HierarchicalBayesianMeasurementModels}}

In Section \ref{sec:measurementModels}, we introduce hierarchical Bayesian measurement models and discuss general methodological considerations.
We go on to discuss pitfalls of correlated measurement error in the simple case of linear regression, where label bias can be studied analytically (see Section \ref{sec:labbias-lin-reg}). After presenting our proposed methodology, we demonstrate the use of Bayesian measurement models in two applications. In Section \ref{sec:example}, we study the simulated criminal justice model considered in \cite{Zanger-Tishler_2023}. 
Next, in Section \ref{sec:diabetes} we consider the problem of predicting diabetes risk based on diagnosis information. We use public health research on diabetes prevalence to adjust for the fact that among diabetics, diagnoses are more likely to be made in those with healthcare access. By adjusting predictions for healthcare status, we achieve more accurate and equitable predictions than possible with regression on the proxy labels. While the examples are chosen to resemble real-world applications, they are not supposed to be case studies. Rather, they are chosen to showcase how our proposed methodology--- Bayesian measurement models---might improve on classical techniques dealing with label bias.

\section{Measurement models for label bias}
\label{sec:measurementModels}
\begin{figure}[tb]
    \centering
\includegraphics[width=\textwidth]{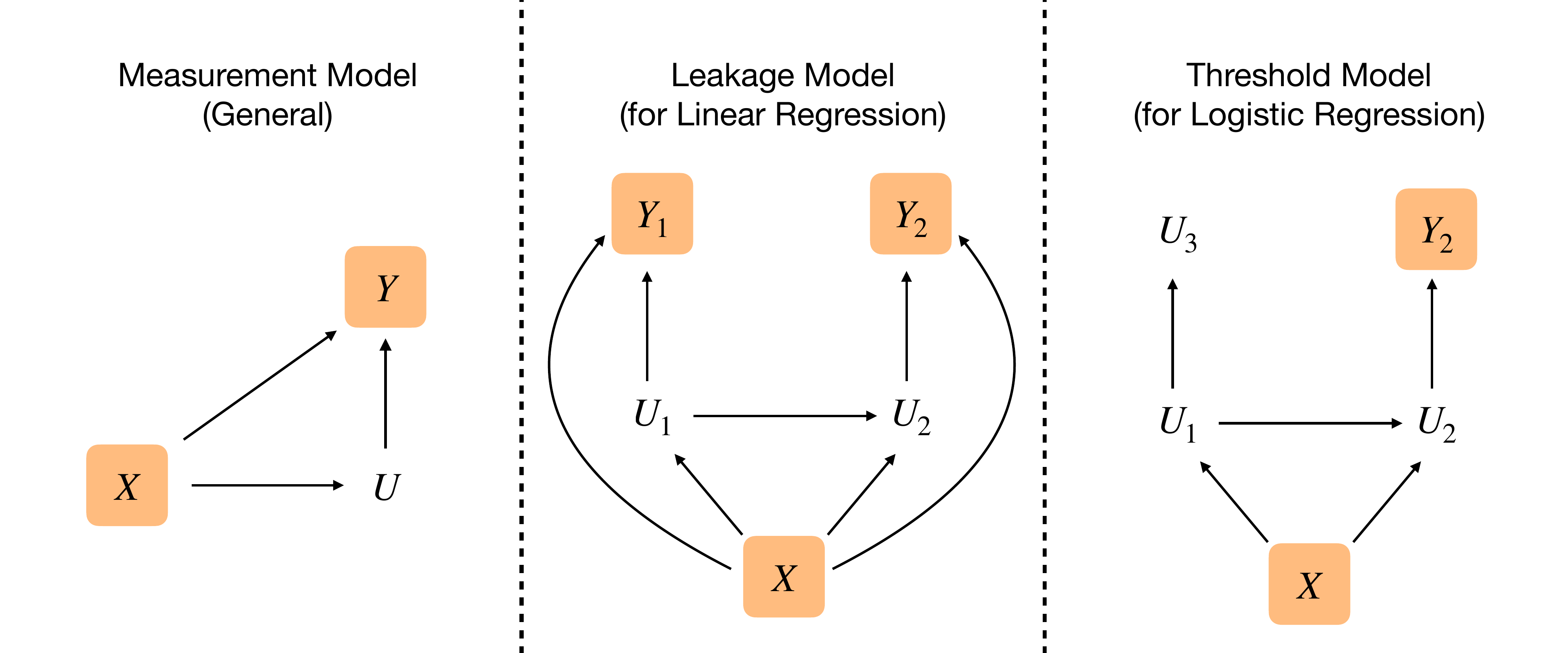}
    \caption{Some measurement models for label bias}
    \label{fig:measurementModels}
\end{figure}
In situations in which only a noisy proxy $y$ of the desired outcome of interest $u$ is available, some model, explicit or implicit, of the measurement process is necessary for accurate and reliable prediction. The classical approach of using regression $\mathbb{E}(y|X)$ on the proxy labels to predict the true outcome $u$ implicitly equates the outcome of interest and the observed proxy outcome. In the case of linear regression, this yields accurate inference if the measurement error is mean independent of the covariates $X$, see Section \ref{sec:labbias-lin-reg}. 
Often there is good reason to believe this is not the case. Measurement models in general, and Bayesian hierarchical models in particular, are a useful tool to model more complicated measuring processes and account for noise that is not independent of the covariates.

The general idea behind measurement models (see Figure \ref{fig:measurementModels}) is to introduce the true outcome $u$ as a latent (unobserved) quantity. Prior knowledge about the application is then used to model the relationship between the covariates $X$, the latent outcomes of interest $u$, and the observed proxies $y$. Because parts of the variables remain unobserved, some of the model parameters are not (or only partially) identified \citep{Gustafson_2015}. Measurement models thus rely on domain knowledge in two ways: The measurement process has to be sufficiently understood to supply a model structure (which includes distributional assumptions about the latent outcomes) as well as reasonable values of the non-identified parameters of the model. We give guidance on how to determine which parameters require strong priors in Appendix \ref{App:DetParamsStrongPri}.

For the identified part of the model, classical advice about Bayesian workflow \citep{gelman2020bayesianworkflow} applies. In particular, posterior predictive checks \citep{Rubin_1984,Gelman_Meng_Stern_1996} can be used to asses model fit.
If parametric assumptions are too rigid, 
 non-parametric components (such as Gaussian processes or splines) can be used.
Another way of adding flexibility and moving beyond the limitations of parametric models is to add unit-specific error terms (such as in the threshold model of Section \ref{Sec:thresholdModel}). 

When only limited prior knowledge is available, non-identified parameters should be treated as sensitivity parameters in a sensitivity analysis \citep{Richardson_Evans_Robins_2011}.
Such an analysis is performed in Section \ref{Sec:Miss}, which details the impact of misspecification of the parameters in a stylized example where the data-generating process is known.  \citet{Gelman_Hennig_2017} discuss the use of informative priors in Bayesian practice more generally and the value of transparency in scientific endeavors.

Measurement models are flexible and can be tailored to the application of interest. 
Here we present two models, a leakage model for linear regression, which we will use to model a stylized example of arrests and crime (see Section \ref{sec:example}), and a threshold model for logistic regression, which we will apply to estimate diabetes risk based on diagnosis data (see Section \ref{sec:diabetes}).
Before we do so, we will illustrate the pitfalls of dependent label bias explicitly in the case of linear regression.
\subsection{Simple illustration: Label bias in linear regression}
\label{sec:labbias-lin-reg}
\begin{figure}[tb]
    \centering
\includegraphics[width=\textwidth]{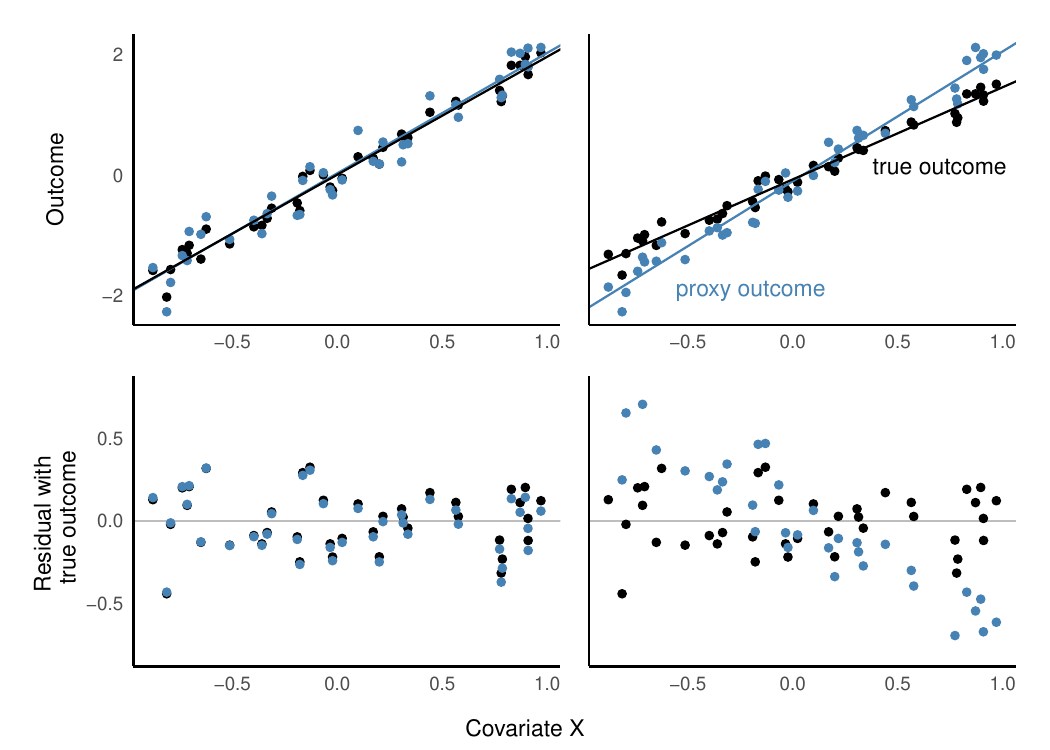}
    \caption{Illustration of label bias in linear regression. (Left) If the measurement errors are uncorrelated with the covariate, regression yields unbiased and consistent estimates. (Right) In the case of dependent measurement errors, regression estimates are biased and inconsistent. Prediction accuracy is degraded.}
    \label{fig:lin-reg-ill}
\end{figure}
In this section, we use the simple case of linear regression to analytically demonstrate 
issues that can arise when using regression on proxy outcomes to predict true outcomes. The validity of this classical approach rests on the assumption that the measurement error is uncorrelated with the covariates. We demonstrate that if this assumption is inaccurate, predictions can be systematically inaccurate. 
Throughout this section, we treat the covariates $X$ as random \citep{Buja_Berk_Brown_George_Pitkin_Traskin_Zhao_Zhang_2016,Buja_Brown_Kuchibhotla_Berk_George_Zhao_2019,Rosset_Tibshirani_2020} allowing them to be correlated with the measurement errors.

We provide three main formulas. First, in Proposition \ref{prop1}, we 
provide a formula for the error of the linear regression solution when fitting on 
a proxy as opposed to the true outcome. While our proposition is focused on regression on proxies, it is similar to the well-known omitted variable bias \citep{Wooldridge}.
Proposition \ref{prop2} demonstrates
that when the proxy is correlated with predictors, then the prediction error is also 
correlated with the predictors. Finally, Proposition \ref{Thm:LabelBias-Is-Bad} provides a lower bound on the prediction error when using a proxy outcome in terms of the prediction 
error when using the true outcome. The primary purpose of these propositions is to demonstrate the systematic errors that can arise when using proxy labels in a simple setting that can be studied analytically. Proofs can be found in Appendix \ref{App:Proof}.

We start by assuming that some true outcome, $u$, and a proxy outcome, $y$, are $n$-dimensional random vectors. We also assume that $X$ is an $n\times m$ random matrix of centered covariates with a leading column of ones such that $\E(X^\top X)$ is full rank, i.e., the covariates are not multicollinear. 
We assume $(X, u, y) \sim P $ where $P$ is some probability distribution over the covariates, true outcome, and proxy outcome.\footnote{We assume that the expectations taken with respect to $P$ in the proofs of Section \ref{App:Proof} all exist.}
We define $\beta$ to be the expected solution to linear regression with covariates $X$ and data $u$. That is,  
\begin{align}\label{beta_def}
\beta = \argmin_{w} \,\, \E(\|X w - u \|^2).
\end{align}
The expected solution to the linear regression changes when using the proxy outcome $y$ and the same covariates $X$. The expected solution with a proxy outcome is given by the following proposition.
\begin{proposition}[Proxy outcome regression solution]
Let $(X, u, y) \sim P $. Then, 
\begin{align}\label{beta_proxy}
\argmin_{w} \,\, \E(\|X w - y \|^2) = (1 + \gamma) \beta + \alpha
\end{align}
where the vector $[\alpha \,\, \gamma] \in \R^{m + 1}$ is the expected solution to the linear regression with outcome $u - y$ (the measurement error) and $n \times (m + 1)$ matrix of covariates $[X \,\, u]$. That is, 
\begin{align}\label{alpha_def}
[\alpha \,\, \gamma] = \argmin_{w} \,\, \E(\| M w - e \|^2) 
\end{align}
where $e$ is the measurement error defined by $e = u - y$ and where $M$ is 
defined to be the $n \times (m + 1)$ random matrix $[X \,\, u]$. 
\label{prop1}
\end{proposition}
That is, if the measurement error $e$ is uncorrelated with the 
covariates and the outcome, then, in expectation, $\beta$ is recovered from the proxies.
On the other hand, correlation between the measurement error
and the covariates or the outcome will introduce error in the approximation
of $\beta$. That error, $ \gamma \beta + \alpha$, is obtained from 
combining \eqref{beta_def} and \eqref{beta_proxy}.
This can pose problems in causal investigations \citep{Knox_Lucas_Cho_2022}
and even in predictive settings.
The right panels of Figure \ref{fig:lin-reg-ill} provide an illustrative example of 
error introduced by the use of a proxy outcome in the linear regression setting. 
We demonstrate the case where $m = 2$, i.e., $X$ consists of an intercept and one 
predictor. 

When label bias introduces error into the solution to a linear regression, 
the predictions made using that linear regression will be systematically distorted. We define the predictions as $\hat{\mathbb{E}}(y|X) = X (X^\top X ) ^{-1} X^\top y$.
The following proposition provides a formula for the covariance between 
the covariates, $X$, and prediction error, $u -  \hat{\mathbb{E}}(y|X)$.
\begin{proposition}[Covariance of covariates and prediction error]
Let $(X, u, y) \sim P $. Then, 
\begin{align}
\E[(u-\hat{\mathbb{E}}(y|X))^\top X] 
     \, &= \, -(\gamma \beta +\alpha)^\top \E(X^\top X)
\end{align}
where $\beta$ is defined in \eqref{beta_def}, and $\alpha, \gamma$ are defined
in \eqref{alpha_def}. 
\label{prop2}
\end{proposition}
That is, if there is correlation between the covariates and the measurement 
error $u - y$, then the prediction error 
will also be correlated with the covariates. 
This shows that the use of proxy labels may introduce \textit{systematic} disparities in predictions. These disparities are liable to negatively affect downstream decisions based on them \citep{Van_Calster_Vickers_2015,Parastouei_Sepandi_Eskandari_2021} and may lead to protected groups being underserved, thus violating certain conceptions of algorithmic fairness\citep{Dwork_Hardt_Pitassi_Reingold_Zemel_2011,Hardt_Price_Price_Srebro_2016,JMLR:v24:22-1511}.

In our last proposition, we compare prediction error when fitting with the true outcome
to prediction error when using a proxy. 
In particular, we provide a lower bound for the mean squared error (MSE) in the true 
outcome using linear regression predictions trained on a proxy in terms of the 
MSE in the true outcome using linear regression trained on the true outcome. 
We show that label bias degrades prediction accuracy when using linear regression because of the systematic disparities in prediction caused by the correlation between the measurement error and the outcome $u$ and covariates $X$.
\begin{proposition} [Prediction error with true outcome versus proxy]
Let $(X, u, y) \sim P $. Then we have
\begin{align*}
    \mathrm{MSE}(u,\hat{\mathbb{E}}(y|X)) \geq \mathrm{MSE}(u,\hat{\mathbb{E}}(u|X)) + (\gamma \beta +\alpha)^\top \E(X^\top X)(\gamma \beta +\alpha)
\end{align*}
\label{Thm:LabelBias-Is-Bad}
where $\beta$ is defined in \eqref{beta_def}, and $\alpha, \gamma$ are defined in \eqref{alpha_def}.
\end{proposition}
In Section \ref{sec:example} and \ref{sec:diabetes}, we will see that given sufficient domain knowledge these systematic disparities in prediction can be mitigated, improving both overall prediction accuracy and reducing the risk of exacerbating disparate outcomes of downstream decisions.

\subsection{Leakage model for linear regression}
\label{Sec:leakageModel}
Measurement models are tailored to specific applications and depend on both knowledge about the structure and the parameters of the measurement process. In this section, we describe a \textit{leakage model} for linear regression based on the stylized criminal justice example we will study in Section \ref{sec:example}.
Suppose we observe a proxy label $y_t$ at two different time points $t \in \{1,2\}$.
These proxies depend both on the observed covariates $X$ and the true outcomes $u_t$. 
In the criminal justice example, arrests are proxies $y_t$ for the true outcome $u_t$ of crime. 
Not all crime leads to arrests, so there is a degree of leakage between proxies and latent outcomes of interest. We assume that the proxies do not influence each other; that is, the entire temporal relationships in the model are driven by the dependence of $u_2$ on $u_1$.\footnote{The center panel of Figure \ref{fig:measurementModels} has an arrow from $u_1$ to $u_2$, implying a causal relationship if the figure is understood as a directed acyclic graph. Our model, however, does not rest on this assumption and is still applicable if $u_1$ and $u_2$ are just assumed to be correlated.} We are interested in learning this relationship and inferring $u_t$ based on $y_t$ and $X$. This assumption is based on our knowledge of the data-generating process for the example we are studying here (see Figure \ref{Fig:SEM}). In other situations, we might assume that the proxies at time $t=1$ influence the outcome at time $t=2$, for example, arrests might deter future crime. Measurement models are flexible enough to allow for this and our model is easily extended to this case. 

This situation studied here is illustrated in the center panel of Figure \ref{fig:measurementModels} and can be modeled by the following Bayesian hierarchical model:
\begin{align} \nonumber
\label{Eq:leakage}
    y_1 | u_1, \alpha,\gamma,\sigma_y \, &\sim \, \text{normal}(X \alpha + \gamma u_1, \sigma_y) 
    \\[2pt] 
     y_2 | u_2, \alpha,\gamma,\sigma_y \, &\sim \, \text{normal}(X \alpha + \gamma u_2, \sigma_y) 
     \\[2pt] \nonumber
      u_1 | \beta \sigma_u\, &\sim \, \text{normal}(X \beta , \sigma_u) 
      \\[2pt] \nonumber
      u_2 | u_1, \beta, \eta, \sigma_u\, &\sim \, \text{normal}(X \beta + \eta(u_1 - X\beta) , \sigma_u \sqrt{1-\eta^2}),
\end{align}
with appropriate priors on all parameters. Because the true outcomes $(u_1,u_2)$ remain unobserved, this model is only partially identified \citep{Gustafson_2015}.
We give guidance on identifying parameters that require strong priors in Appendix \ref{App:DetParamsStrongPri}. 
In this example, weak priors suffice for ($\alpha$, $\eta$,$\sigma_y$) when we use  strong priors on ($\beta$,$\gamma$, $\sigma_u$). We will use this model for a stylized example of criminal behavior and arrests in Section \ref{sec:example}.
\subsection{Threshold model for logistic regression}
\label{Sec:thresholdModel}
Here we develop a \textit{threshold model} for logistic regression. We deploy this model for diabetes prediction in Section \ref{sec:diabetes}.

Suppose we observe binary proxy labels $y \in \{0,1\}$ instead of a binary outcome of interest $u_3$. The proxies are indicative of the true outcome but they are not fully reliable, that is there are cases of $u_3$ that $y$ does not indicate. In our diabetes example, $u_3$ indicates diabetes disease status. Not everyone with diabetes is diagnosed, however, so diagnosis $y$ is not a fully reliable proxy. 

This situation can be modeled by introducing two (continuous) latent characteristics $u_1$ and $u_2$ that cause $u_3$ and $y$, respectively, by crossing a threshold:
\begin{align} \nonumber
y \, &= \, \begin{cases}
    1 \, \, \text{if}\, \, u_2 \geq 0 \\
    0 \, \, \text{else}
\end{cases} \\[2pt] \nonumber
    u_1 | \beta \, &\sim \, \text{logistic}(X\beta,1) \\[2pt] 
    u_2 \, &= \,  u_1 - t(X) - e \\[2pt] \nonumber
    u_3 \, &= \, \begin{cases}
    1 \, \, \text{if}\, \, u_1 \geq 0 
    \\[2pt] \nonumber
    0 \, \, \text{else}
\end{cases} \\[2pt] \nonumber
e \, &\sim \, \text{normal}^+(0,0.1).
\end{align}
The thresholds $t(X)$ can depend on covariates $X$ allowing for disparities in how accurate the proxies are for different subpopulations.
The structure of the model is illustrated in the right panel of Figure \ref{fig:measurementModels}.

In the diabetes example of Section \ref{sec:diabetes}, the latent variable $u_1$ can be understood as quantifying the severity of diabetes. We assume that for uninsured people symptoms have to be more severe to be diagnosed. This is modeled by introducing insurance-dependent thresholds $t(\text{health insurance})$ that offset the latent characteristic $u_2$ that determines diagnosis. 
By introducing $e>0$, we allow for idiosyncratic behavior that impacts the proxy, e.g. patient's personal propensity to visit a doctor.

We assume that there are no false positives, that is, $u_3 \geq y$. In essence, this assumes that people are not mistakenly diagnosed with diabetes and that their response about their diagnosis is truthful. If we have reasons to believe this to be false, we could allow $t(\text{health insurance})$ to be random, leading to false positive diagnoses for a fraction of the population.

The latent characteristic $u_1$ depends linearly on the covariates, so the threshold model closely resembles ordinary logistic regression \citep{Gelman_Carlin_Stern_Dunson_Vehtari_Rubin} but allows for discrepancies between the outcomes of interest $u_3$ and the observed labels $y$.

In Section \ref{sec:diabetes}, we use this model to predict diabetes risk based on diabetes diagnosis with varying thresholds based on health insurance status.

\section{Stylized example: Criminal behavior and arrests}\label{sec:example}
\begin{figure}[tb]
\centering
\includegraphics[width=0.49\textwidth]{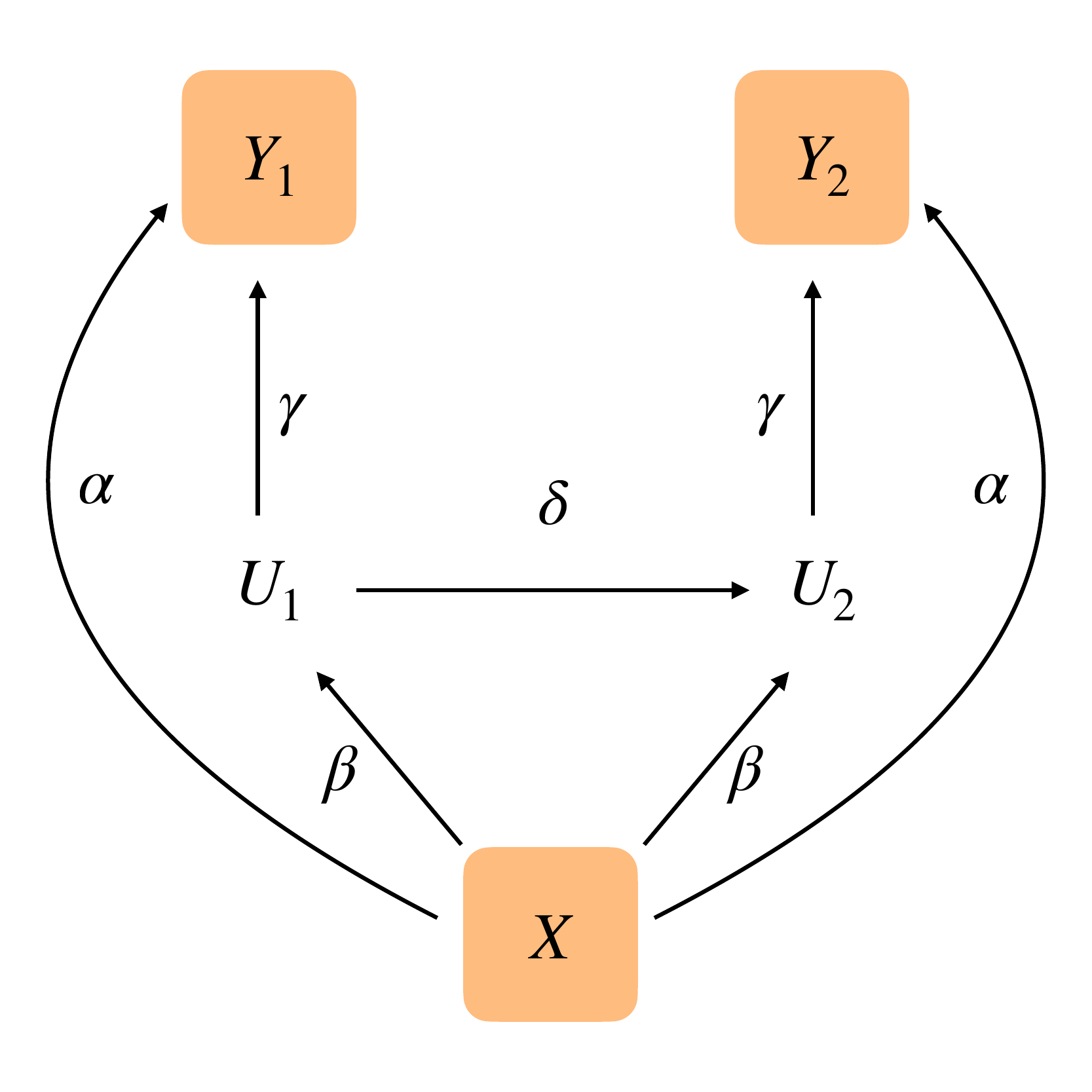}
\caption{Data-generating process for \citet{Zanger-Tishler_2023} stylized example of criminal behavior (true outcome) and arrest (proxy outcome). Observed variables are in orange.}
\label{Fig:SEM}
\end{figure}

Figure \ref{Fig:SEM} portrays the data-generating process for a stylized example of label bias in \citep{Zanger-Tishler_2023}.
This model simulates individual-level behavior ($u_0$ and $u_1$) and arrest outcomes ($y_0$ and $y_1$) at two time points. Arrests depend both on an individual's behavior and the individual's neighborhood ($X$). 
This is a linear structural equation model,
\begin{align} \nonumber
\label{Eq:SEM}
    X &\sim \text{normal}(0,\sigma_X) \\
\left.  \begin{bmatrix}
u_0  \\
u_1   
\end{bmatrix} \right| X  &\sim \text{MVN}\bigg(\begin{bmatrix}
\beta  X  \\
\beta X   
\end{bmatrix},\begin{bmatrix}
\sigma_u^2 & \delta \\
\delta & \sigma_u^2 
\end{bmatrix}\bigg) \\\nonumber
y_0 |X,u_0& \sim \text{normal}(\alpha X + \gamma u_0, \sigma_y) \\\nonumber
y_1 |X,u_1& \sim \text{normal}(\alpha X + \gamma u_1, \sigma_y).
\end{align} 
\citet{Zanger-Tishler_2023} show how to set the variances of the exogenous variables such that the remaining variables ($X$, $u_0$, $u_1$, $y_0$, and $y_1$) are standardized and can be interpreted as the extent to which an individual differs from the population average. For example, $u_0$ is interpreted as how criminal an individual is compared to the population, and $X$ as the level of police enforcement in a neighborhood. 

The label bias in this problem arises because only arrests ($y_0$ and $y_1$) and neighborhood ($X$) are observable. Criminal behavior, $(u_0,u_1)$, which is the true outcome of interest, is not observable and therefore arrests are used as a proxy for criminal behavior. We have two regression models, a simple one $\mathbb{E}(y_1|y_0)$ and a complex one $\mathbb{E}(y_1|y_0,X)$.
\citet{Zanger-Tishler_2023} show (see Corollary 1) that it is preferable (in terms of expected squared difference between true and predicted outcome) to not include an additional feature if the correlation of that feature with the true and proxy outcome conditional on other covariates have differing signs. For the stylized example here, they show that this is the case for the inclusion of neighborhood, $X$, in a model for predicting criminal behavior based on arrests $y_0$ when the correlation between neighborhood and criminal behavior, $\beta$, is small.

 This simple example illustrates the theoretical insight of \citet{Zanger-Tishler_2023}, that the inclusion of additional features can degrade the predictive accuracy of regression models in the presence of label bias.

\subsection{Bayesian measurement model}
\begin{figure}[tb]
\centering
\includegraphics[width=0.7\textwidth]{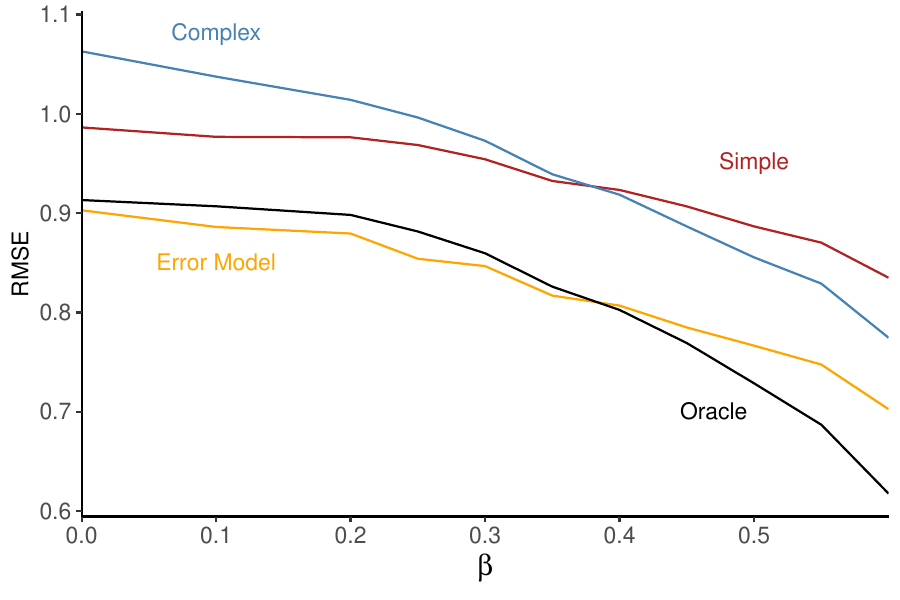}
\caption{Root mean squared error for simple and complex regression models trained on proxy outcomes in comparison with a Bayesian error model. The prediction accuracy of the error model is superior to both regression models for all $\beta$ and comparable to an oracle model (trained on the true outcomes). }
\label{Fig:errorModel}
\end{figure}
Here we propose to use the leakage model for regression introduced in Section \ref{Sec:leakageModel}, which improves on both the simple and complex estimators of Section \ref{sec:example}. In this stylized example, the data-generating process is known (see Figure \ref{Fig:SEM}) and corresponds to the model structure of our measurement model; compare Equations (\ref{Eq:leakage}) and (\ref{Eq:SEM}).
Making an informed decision about the model structure is the first step when using a measurement model. The second step is the choice of priors. While for some parameters weak priors are sufficient, the model is only partially identified \citep{Gustafson_2015} necessitating strong priors on the non-identified part of the model (see Appendix \ref{App:DetParamsStrongPri}).\footnote{We assume $\sigma_u$ is known.}
\begin{align*}
    \sigma_y &\sim \text{normal}^+(0,1) \\
    \alpha &\sim \text{normal}(0,1) \\
    \eta &\sim \text{normal}(0,0.2) \\
    \beta &\sim \text{normal}(\beta_{true},0.1) \\
    \gamma &\sim \text{normal}(\gamma_{true},0.1).
\end{align*}
The strong priors are informed by our knowledge of the data-generating process. 
In general, when prior knowledge is limited, the non-identified parameters of the model should be treated as sensitivity parameters in a sensitivity analysis.
Section \ref{Sec:Miss} performs such a sensitivity analysis and investigates the impact of misspecification of the nonidentifiable parameters.

Here we have implicitly switched from a prediction setting, in which we are only interested in $\mathbb{E}(u_1|y_0,X)$, to an inference setting where we are modeling the joint distribution $P(u_0,u_1,y_0,y_1)$. If we require predictions on a set of new variables for which only the features are observed, we can do so by incorporating the unobserved outcomes as missing variables.

Figure \ref{Fig:errorModel} shows that the error model performs better than both of the regression models with its performance being comparable with a regression on the true labels. 

In practice, the measurement process might be more complicated than in this stylized example. For example, we might have a multitude of correlated covariates each impacting the measurement error to varying degrees. Our approach is flexible enough to cover such cases, however, with increasing complexity, it may become less likely that sufficient domain knowledge is available to tightly constrain all non-identifiable parameters. \citet{Gelman_Hennig_2017} discuss the value of transparency and the use of informative priors in practice more generally. 
While this may limit the efficacy of our method, it also limits the applicability of classical methods. If the measurement process cannot be accounted for, prediction accuracy can be arbitrarily degraded (see Proposition \ref{Thm:LabelBias-Is-Bad}). Domain knowledge is crucial for predictions with label bias. If the structure of the measurement process is known but the values of its non-identifiable parameters are not, our method offers two advantages over classical methods: For one, using wide priors, we can propagate our uncertainty about the measurement process to the predictions. Secondly, we can vary these parameters systematically to check the sensitivity of the predictions to them; see Section \ref{Sec:Miss}. Neither of these can easily be done for classical methods, risking practitioners to be overly confident in their predictions under label bias.

\subsection{Disparate predictions}
\begin{figure}[tb]
\centering
\includegraphics[width=0.7\textwidth]{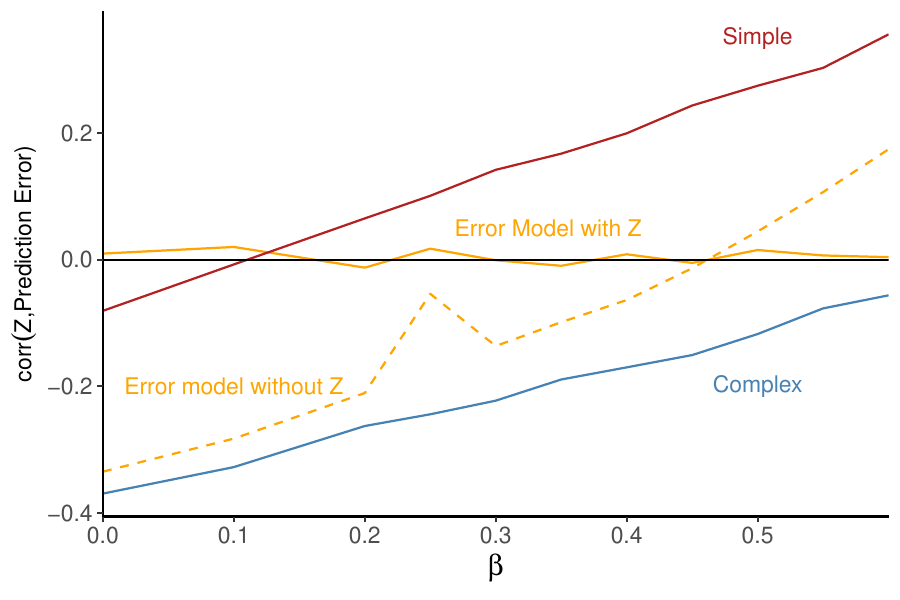}
\caption{Correlation between prediction error $u_1 - \hat u_1$ and neighborhood police enforcement $X$ for simple and complex regression models trained on proxy outcomes in comparison with a Bayesian measurement model. While both regression models produce predictions that are systematically biased based on neighborhood, the measurement model has prediction errors uncorrelated with neighborhood. The dashed line shows that without using neighborhood information, the measurement model also produces biased predictions.}
\label{Fig:corrSem}
\end{figure}
In many applications systematic disparities in prediction between different subgroups can negatively affect downstream decisions \citep{Van_Calster_Vickers_2015,Parastouei_Sepandi_Eskandari_2021}, and, depending on our decision process, lead to decreased fairness \citep{Dwork_Hardt_Pitassi_Reingold_Zemel_2011,Hardt_Price_Price_Srebro_2016,JMLR:v24:22-1511}.
For both the simple and complex regression models studied in \citep{Zanger-Tishler_2023}, prediction errors are correlated with the degree of policing in a neighborhood $X$, i.e. they systematically under- or overpredict crime rates based on neighborhood. This correlation strongly depends on the relationship between neighborhood and behavior as well as arrests.
On the other hand, prediction errors of the Bayesian measurement model are uncorrelated with neighborhood $X$ as long as the neighborhood is accounted for in the model and priors are specified correctly (see Section \ref{Sec:Miss}). Removing neighborhood information from the model slightly decreases predictive performance but introduces dependence between prediction errors and neighborhood $X$. We plot correlations between prediction errors and neighborhood in Figure \ref{Fig:corrSem}.

This shows that modeling the measurement process is key for both overall accurate predictions as well as minimizing systematic disparities in prediction.
 
\subsection{The impact of misspecification}
\label{Sec:Miss}
\begin{figure}[!tb]
\includegraphics[width=\textwidth]{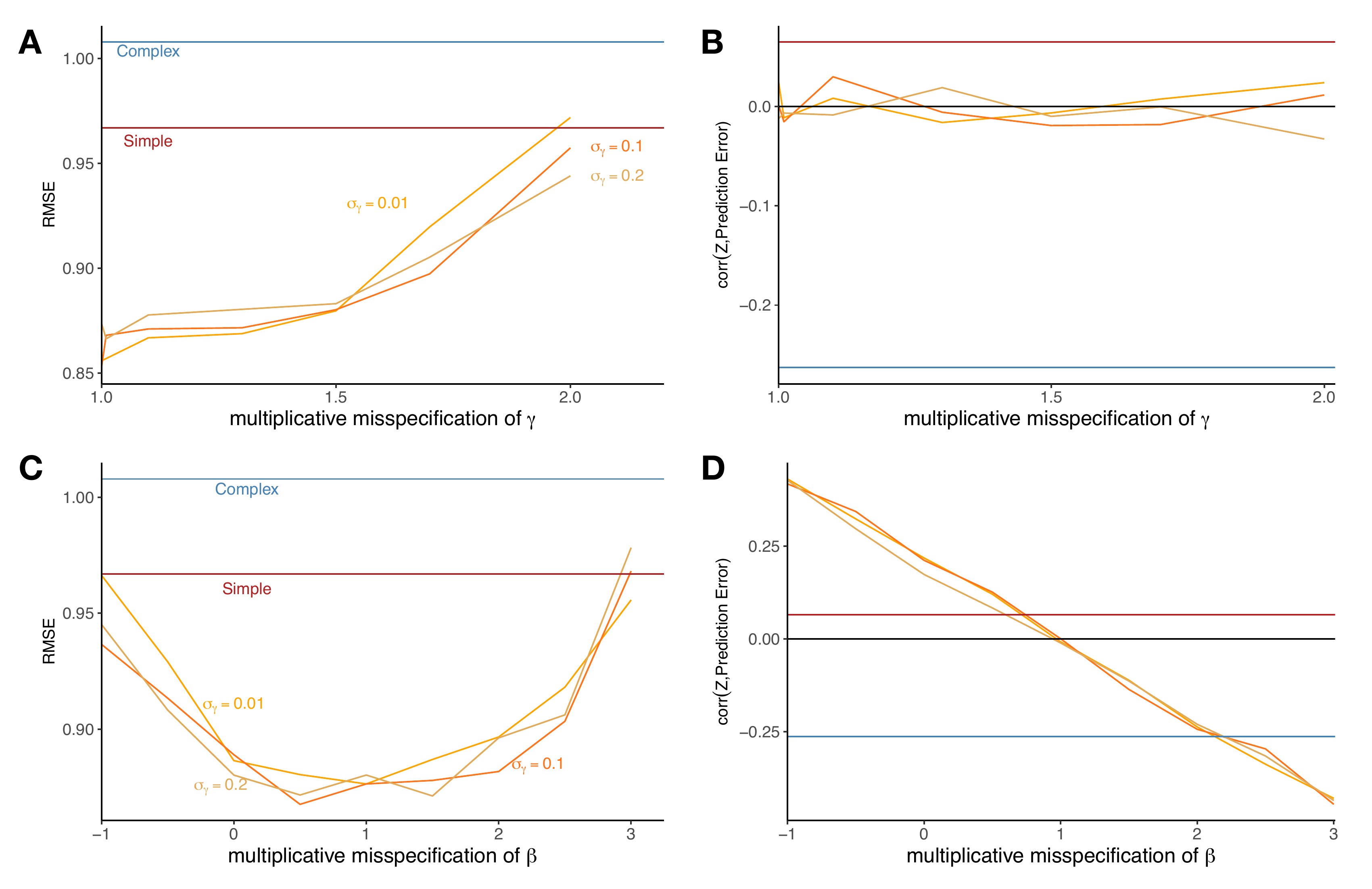}
  \caption{Impact of multiplicative misspecification ($m \times$ true parameter) of $\gamma$ (upper row, $\gamma_{\text{true}}=0.4$) and $\beta$ (lower row, $\beta_{\text{true}}=0.2$) on predictive performance and disparity in prediction accuracy with respect to neighborhood police enforcement. }
  \label{Fig:misscalibration}
\end{figure}

The reliability and accuracy of predictions based on proxies crucially depend on the validity of assumptions we make about the measurement process. 
In the previous section, we have explored the benefits of measurement models when we can correctly account for the measurement process. While there are real-world examples in which this can be done (see Section \ref{sec:diabetes}), this may be unrealistic in the case of predicting crime rates. Despite various proxies being available (for example data on self-reported criminal offending \citep{NLSY97}), the true crime rate is empirically inaccessible. 
Predictions of the true crime rates thus hinge on untestable assumptions that are often obscured by being stated only implicitly, as is often the the case when using regression trained on proxy labels. 

Measurement models, however, force us to make our assumptions transparent and allow to test the prediction's sensitivity to them. 
Figure \ref{Fig:misscalibration} shows the impact of misspecifying the (non-identifiable) parameters $\beta$ and $\gamma$ in our measurement model (\ref{Eq:leakage}). While misspecification of either leads to degraded prediction accuracy, correctly specifying $\beta$---the relationship between neighborhood policing and criminal behavior---is paramount to mitigate systematic disparities in prediction.
In the case of regression models that use proxies to predict crime, these assumptions are often only implicit (and cannot be easily varied), with \citet{Zanger-Tishler_2023} criterion being a step in the direction of transparency. Assumptions being made only implicitly, however, neither implies the results to be agnostic or robust with regards to the underlying measurement process. Figure \ref{Fig:errorModel} and \ref{Fig:corrSem} show that the accuracy and systematic disparities in prediction of both the simple and complex model (as well as the decision which one to choose for prediction) depend on the underlying relationship of crime and neighborhood policing as well.

Given that criminal behavior, let alone its relationship with policing, is virtually impossible to quantify, predicting crime based on arrests is always skewed by our prior assumptions about crime \citep{Biderman_Reiss_1967,Hinton_2016}.
\section{Empirical example: Health insurance and diabetes}
\label{sec:diabetes}
\begin{figure}
    \centering
    \includegraphics[width=\textwidth]{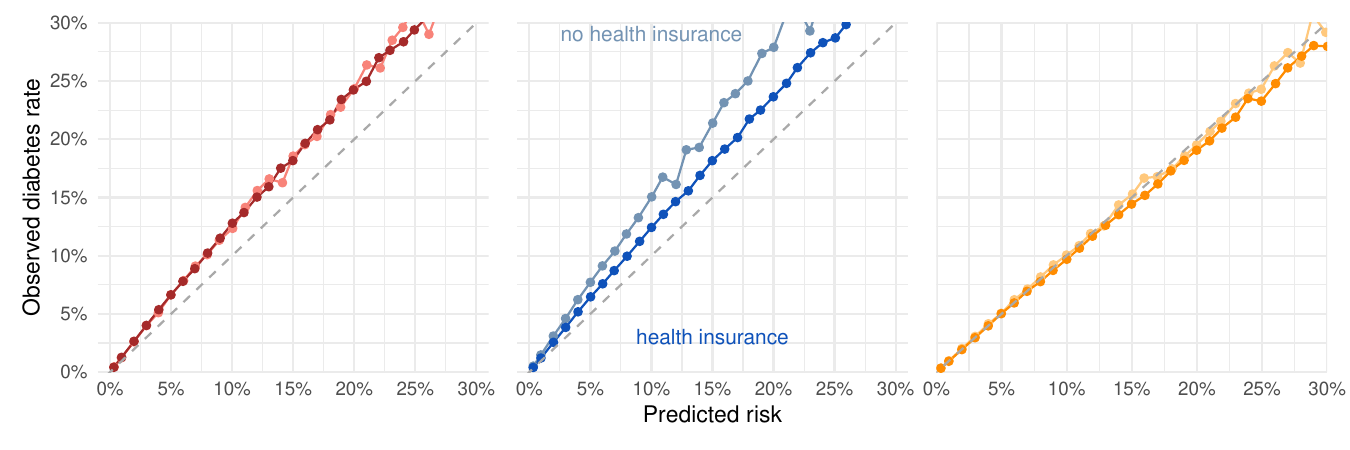}
    \caption{Predicted diabetes risk against diabetes rate against observed diabetes rate estimated with logistic regression on the true outcomes for a simple (red), complex (blue), and measurement model (orange) by health insurance status (darker hue: insured, lighter hue: uninsured). For both insured and uninsured people, our measurement model performs better than both regression models closely matching predictions of logistic regression on the true outcomes (dashed gray line).}
    \label{fig:diabetes}
\end{figure}
It is estimated that more than $10\%$ of the U.S. population has some form of diabetes \citep{CDC_diab}. 
While early identification of diabetes is crucial as behavioral counseling, dietary interventions, increased physical activity, or pharmacologic therapy may improve future health outcome \citep{TaskForce}, testing for diabetes also comes with monetary and personal costs. In practice, this necessitates risk-based screening decisions \citep{Duan_Kengne_Echouffo-Tcheugui_2021}.
In the case that diagnosis information is used to infer the model, predictions will suffer because of label bias. Due to a variety of factors, many people with diabetes have never been diagnosed, making diagnosis an imperfect proxy. For example, it has been estimated that $29\%$ of American diabetics without health insurance remain undiagnosed compared to only $16\%$ with some kind of health insurance \citep{Fang_Wang_Coresh_Selvin_2022}, a difference that can easily be explained by impeded access to healthcare services.
Our analysis is based on publicly available data from the National Health and Nutrition Examination Survey \citep{NHANES}, which provides information about both diagnosed (self-reported information of having been diagnosed with diabetes in the past) and undiagnosed diabetes based on measured blood sugar levels. A ready-to-analyze version of this dataset is provided by \citet{Coots_2023}.
This offers an empirically realistic situation of label bias with the necessary ground-truth data to evaluate the advantages and disadvantages of a measurement model, as compared to simple regression or the approach recommended by \citet{Zanger-Tishler_2023} to drop a predictor.

The left and center panels of Figure \ref{fig:diabetes} show that this situation suffers from the phenomenon described in \citet{Zanger-Tishler_2023}: inclusion of information on the insurance status degrades predictive power when using regression on the proxy labels. For both models, label bias leads to underestimation of diabetes risk. If this bias is not taken into account, decisions based on an optimal treatment threshold are liable to be harmful \citep{Rothblum_Yona_2023} because people who would benefit from treatment will not receive it. 
When insurance status is included as a covariate, disparate predictions are prone to lead to decisions that further under-serve the uninsured population, violating conceptions of algorithmic fairness \citep{Dwork_Hardt_Pitassi_Reingold_Zemel_2011,Hardt_Price_Price_Srebro_2016,JMLR:v24:22-1511}.

We model this situation with the measurement model presented in Section \ref{Sec:thresholdModel}. Here, $y$ are binary indicator for diabetes diagnosis (proxy labels) and $u_3$ indicates diabetes (true outcomes, assumed to not be observed). $u_1$ is a latent variable that can be understood as the underlying severity of diabetes. We assume that for uninsured people, the severity of symptoms has to be higher to be diagnosed. To account for that, we introduce health insurance dependent thresholds $t(\text{health insurance})$ that offset the latent characteristic $u_2$ that determines if a patient is diagnosed. 

This measurement model critically depends on the thresholds $t(\text{health insurance})$, which cannot be inferred from diagnosis data alone. We can, however, use prior knowledge, as in \citep{Fang_Wang_Coresh_Selvin_2022}, to inform our choice. In Appendix \ref{App:threshold}, we discuss in detail how we determine the thresholds. The right panel of Figure \ref{fig:diabetes} shows that this measurement model based on diabetes diagnosis correcting for impeded access to health care services is well calibrated and predicts diabetes risk better than either a simple or complex regression model. Table \ref{tab:app_diabetes} in Appendix \ref{App:Diabetes} shows improved prediction quality across a range of metrics for classification.

\section{Conclusion and discussion}
The use of imperfect proxies as dependent variable is ubiquitous in quantitative research in the social sciences.  These analyses suffer from label bias, which is often assumed to be a minor problem. If the measurement error is correlated with covariates, label bias can have detrimental effects even in purely predictive settings. In these situations, predictions will suffer from systematic disparities---that ,is, we will over- or underpredict the outcome systematically based on the covariates. If the measurement errors are correlated with membership in a protected group, these systematic disparities in prediction will not only lead to degraded prediction accuracy but may also be a concern from an algorithmic fairness perspective. In our diabetes example, see Section \ref{sec:diabetes}, label bias leads classical predictions of the diabetes risk to systematically underpredict true risk, and more so for uninsured people. Decisions based on these estimates will consequently under-serve uninsured people.

In this paper we advocate the use of Bayesian measurement models to mitigate these problems. We show that measurement models are preferable to classical regression models in  two examples:
a stylized criminal justice example, in which the data-generating process is known (see Section \ref{sec:example}), and a real-world example where we estimate diabetes risk based on diagnosis information (see Section \ref{sec:diabetes}). 

We find that when sufficient knowledge about the measurement process is available, these models can mitigate systematic disparities in prediction allowing for more accurate and fairer downstream decisions. 
Our method explicitly requires the user to model the measurement process. This highlights the importance of assumptions about the relationship between measurement error with covariates for reliable, equitable, and accurate predictions. While these assumptions often remain implicit in classical regression methods, our measurement model helps users to make them more transparent. With this transparency also comes the benefit of being able to test the sensitivity of the predictions to the assumed measurement process. This kind of sensitivity analysis is not easily available for classical methods. Overall, this can allow users to better question if enough domain knowledge is at hand to judge if the proxies are useful and to ensure the fairness of downstream decisions based on them.

While we advocate for modeling the measurement process to mitigate systematic disparities in prediction to achieve fairer downstream decisions, we need to firmly state that this cannot be taken as general advice.
Using information necessary in the modeling of proxies, such as protected class status, may be in itself problematic and violate the legal doctrine of “no disparate treatment” (for example the Equal Protection Clause of the U.S. Constitution’s Fourteenth Amendment). This is a fundamental tension and cannot be resolved in general. Any application based on data that is skewed by societal injustices will require careful political, social, and legal consideration. Our paper should, however, be a general warning against the uncritical uses of classical regression methods when faced with this kind of data: in these situations, predictions can suffer from systematic disparities, and decisions based on them can further exacerbate the social injustice that skewed the data.

\section*{Acknowledgments}
We thank Sharad Goel and Michael Zanger-Tishler for their recommendation to use the diabetes example and helpful discussions about results. We also extend our gratitude to Reviewer 2 for their thorough feedback including on technical details.
We thank the U.S. Office of Naval Research for partial support of this work.

\bibliography{sample}

\pagebreak

\appendix
\section{Proofs of Section \ref{sec:labbias-lin-reg}}\label{App:Proof}
This appendix includes proofs of the propositions of Section \ref{sec:labbias-lin-reg}.
For readability, we restate the propositions before giving their proofs. 

For the remainder of this section we assume that $X$ is a random 
$n \times m$ matrix, and $u, y \in \R^n$ are random vectors such that 
$(X, u, y) \sim P$ where $P$ is some probability distribution such that the 
moments taken throughout this section exist. In this random-X setting \citep{Wooldridge,Buja_Berk_Brown_George_Pitkin_Traskin_Zhao_Zhang_2016,Rosset_Tibshirani_2020}, the classical (fixed-X) orthogonality relations between covariates and residuals $X^\top \varepsilon \neq 0$ do not hold generally. Instead, we have orthogonality in expectation $\mathbb{E}[X^\top \varepsilon ] = 0$, which is usually referred to as weak-sense orthogonality. 

\begin{myprop}{1}
Let $(X, u, y) \sim P$. Then 
\begin{align}\label{prop1_eq}
\argmin_{w} \,\, \E(\|X w - y \|^2) = (1 + \gamma) \beta + \alpha
\end{align}
where the vector $[\alpha \,\, \gamma] \in \R^{m + 1}$ is the expected solution to the linear regression with outcome $u - y$ (the measurement error) and $n \times (m + 1)$ matrix of covariates $[X \,\, u]$. That is, 
\begin{align}\label{alpha_def2}
[\alpha \,\, \gamma] = \argmin_{w} \,\, \E(\| M w - e \|^2) 
\end{align}
where $e$ is the measurement error defined by $e = u - y$ and where $M$ is 
defined to be the $n \times (m + 1)$ random matrix $[X \,\, u]$. 
\end{myprop}

\begin{proof}
We define the $n-$ dimensional random vector $\varepsilon$ by the formula 
\begin{align}\label{eps_def}
\varepsilon =  u - X\beta 
\end{align} and 
observe that
\begin{align}
\E(X^\top \varepsilon) = \mathbf{0}
\end{align}
follows from the combination of \eqref{beta_def} and the \eqref{eps_def}. 
That is, the residual vector $\varepsilon$ is uncorrelated with the covariates 
(columns of $X$). 
Defining $e$ to be the measurement error, $e = u - e$, we 
represent $e$ as a linear combination of the columns of $X$, the 
outcome $u$, and a residual. Specifically, we have 
\begin{align}
\label{eq:lin-proj-error}
    e = X \alpha + \gamma u + r,
\end{align}
where 
\begin{align}
[\alpha \,\, \gamma] = \argmin_{w} \,\, \E(\| M w - e \|^2) 
\end{align}
where $M$ is the $n \times (m + 1)$ matrix $[X \,\, u]$ and $r = e - (X \alpha + \gamma u)$ is uncorrelated with $X$ and $u$. 
Using \eqref{eps_def} and the fact that $e = u - y$, we have 
\begin{align}\label{y_eq}
    y & = X \beta + e + \varepsilon.
\end{align}
Substituting \eqref{eq:lin-proj-error} into \eqref{y_eq} yields,
\begin{align}
\label{eq:simple-proxy-eq}
    y \, &= \, \gamma u + X (\beta+\alpha) + r+ \varepsilon \\ 
    \, &= \, X \big((1+\gamma)\beta + \alpha \big) + \tilde \varepsilon,
\end{align}
where we define $\tilde \varepsilon \equiv (1+\gamma)\varepsilon + r$. 
Equation \eqref{prop1_eq} follows immediately from the fact that $\tilde \varepsilon$ is 
uncorrelated with $X$.
\end{proof}

\begin{myprop}{2}
Let $(X, u, y) \sim P$. Then 
\begin{align}
\E[(u-\hat{\mathbb{E}}(y|X))^\top X] 
     \, &= \, -(\gamma \beta +\alpha)^\top \E(X^\top X)
\end{align}
where $\beta$ is defined in \eqref{beta_def}, and $\alpha, \gamma$ are defined
in \eqref{alpha_def2}. 
\end{myprop}
\begin{proof}
\begin{align}\nonumber
    \E[(u-\hat{\mathbb{E}}(y|X))^\top X] 
     \, &= \, \E(u^\top X) - \E(\hat{\mathbb{E}}(y|X)^\top X) \\ \nonumber
     \, &= \, \beta^\top \E(X^\top X) - \E((X(1+\gamma \beta +\alpha)+\tilde \varepsilon)^\top X) \\
     \, &= \, -(\gamma \beta +\alpha)^\top \E(X^\top X). 
\end{align}
\end{proof}

\begin{myprop}{3}
Let $(X, u, y) \sim P$. Then we have
\begin{align*}
    \mathrm{MSE}(u,\hat{\mathbb{E}}(y|X)) \geq \mathrm{MSE}(u,\hat{\mathbb{E}}(u|X)) + (\gamma \beta +\alpha)^\top \E(X^\top X)(\gamma \beta +\alpha)
\end{align*}
where $\beta$ is defined in \eqref{beta_def}, and $\alpha, \gamma$ are defined
in \eqref{alpha_def2}. 
\end{myprop}

\begin{proof}
We have
\begin{align*}
     \mathrm{MSE}(u,\hat{\mathbb{E}}(y|X)) = \mathbb{E}[\lVert u-\hat{\mathbb{E}}(y|X)\rVert ^2] = 
     \mathbb{E}[\mathbb{E}[\lVert u-\hat{\mathbb{E}}(y|X)\rVert^2|X]].
\end{align*}
We focus on $\mathbb{E}[\lVert u-\hat{\mathbb{E}}(y|X)\rVert^2|X]$ first. 
The predictions with  linear regression are $\hat{\E}(y|X) = X(X^\top X)^{-1}X^\top y =: Hy$ where $H$ is defined to be the matrix $X(X^\top X)^{-1}X^\top$. We have
\begin{align*}
    \E[\lVert u-\hat{\mathbb{E}}(y|X)\rVert^2|X] 
    \, &= \, 
    \E[\lVert u-\hat{\E}(u|X)+\hat{\E}(u|X)-\hat{\E}(y|X)\rVert^2|X] \\
    \, &= \, 
    \E[\lVert u-\hat{\E}(u|X)\rVert^2|X] + \E[\lVert \hat{\E}(u|X)-\hat{\E}(y|X)\rVert^2|X] \\
    \, &+ \,  2\E[(u-\hat{\E}(u|X))^\top(\hat{\E}(u|X)-\hat{\E}(y|X))|X],
\end{align*}
where $\hat{\mathbb{E}}(u|X) = Hu$.
The last term vanishes $\E[((1-H)u)^\top H(u-y)|X]=0$ because $H^\top = H$ and  $(1-H)H = 0$. Defining $\E[\lVert u-\hat{\E}(u|X)\rVert^2|X]:= \text{MSE}_{u|X}$, we have
\begin{align*}
    \E[\lVert u-\hat{\mathbb{E}}(y|X)\rVert^2|X] 
    \,  = \, \text{MSE}_{u|X} + \E[\lVert H(u-y)\rVert^2|X].
\end{align*}
For $a,b \in \R^n$, we have $a^\top b = \Tr(b^\top a)$. Reminding ourselves that $y-u = e = X\alpha +\gamma u+ r = X(\gamma\beta+\alpha) + \gamma \varepsilon + r $, we can rewrite 
\begin{align*}
\E[\lVert H(u-y)\rVert ^2|X] \, &=\,     \E(e^\top He |X) = \Tr\,  H\,  \E(e e^\top |X) \\
\, &=\,
\Tr\,  H\,  \E[(X(\gamma\beta+\alpha) + \gamma \varepsilon + r) (X(\gamma\beta+\alpha) + \gamma \varepsilon + r)^\top |X] \\
\, &=\,
\Tr\,  H\,  \E[(X(\gamma\beta+\alpha) (X(\gamma\beta+\alpha))^\top |X] \\
\, &+ \, 2 \gamma \Tr\,  H\,  \E[(X(\gamma\beta+\alpha) \varepsilon^\top |X] 
\, + \, 2 \Tr\,  H\,  \E[(X(\gamma\beta+\alpha) r^\top |X] \\
\, &+ \,  \Tr\,  H\,  \E[(\gamma \varepsilon + r)(\gamma \varepsilon + r)^\top |X].
\end{align*}
For the first term, we have
\begin{align*}
    \Tr\,  H\,  \E[(X(\gamma\beta+\alpha) (X(\gamma\beta+\alpha))^\top |X] \, & = \,  \Tr \, X (\gamma\beta+\alpha) (\gamma\beta+\alpha)^\top X^\top\\ 
    \,& = \, (\gamma\beta+\alpha)^\top X^\top X (\gamma\beta+\alpha).
\end{align*}
Taking expectations on both sides yields
 \begin{align*} 
    \E[ \Tr\,  H\, \E[(X(\gamma\beta+\alpha) (X(\gamma\beta+\alpha))^\top |X]]\, &= \,(\gamma\beta+\alpha)^\top \E[X^\top X](\gamma\beta+\alpha)\\ \, &=\, | \mathrm{cov}(u-\hat{\mathbb{E}}(y|X)\, ,\,X )(\gamma \beta +\alpha)|.
\end{align*}
By the definition of linear regression,  we have $\E(X^\top \varepsilon) = 0$ and $\E(X^\top r) = 0$, so that $X$ is uncorrelated with both $r$ and $\varepsilon$. Hence, we have
\begin{align*}
    2 & \gamma \Tr\,  H\,  \E[(X(\gamma\beta+\alpha) \varepsilon^\top |X] 
\, + \, 2 \Tr\,  H\,  \E[(X(\gamma\beta+\alpha) r^\top |X] \\ \, &= \, 2 \gamma \Tr\, X(\gamma\beta+\alpha) \E[\varepsilon^\top |X] + 2 \Tr\, X(\gamma\beta+\alpha) \E[r^\top |X] \\
\, &= \, 2 (\gamma\beta+\alpha)^\top (\gamma X^\top \E[\varepsilon|X] +  X^\top \E[r|X]) 
\end{align*}
for the second line. Taking expectations, we get
\begin{align*}
 \E[2 & \gamma \Tr\,  H\,  \E[(X(\gamma\beta+\alpha) \varepsilon^\top |X] 
\, + \, 2 \Tr\,  H\,  \E[(X(\gamma\beta+\alpha) r^\top |X]] \\
\, & = \, 2 (\gamma\beta+\alpha)^\top (\gamma \E[ X^\top \varepsilon] +   \E[X^\top r]) = 0.
\end{align*}
For the last term, we have 
\begin{align*}
     &\Tr\,  H\,  \E[(\gamma \varepsilon + r)(\gamma \varepsilon + r)^\top |X] \geq 0,
\end{align*}
which implies
\begin{align*}
      & \E[ \Tr\,  H\,  \E[(\gamma \varepsilon + r)(\gamma \varepsilon + r)^\top |X]] \geq 0    
\end{align*}
because 
$H$ (a projection matrix) and $\E[(\gamma \varepsilon + r)(\gamma \varepsilon + r)^\top |X]$ are positive semidefinite. The bound follows from the fact that for $A,B \in \R^{n \times n}$ symmetric and positive semi-definite, there exists $Q\in \R^{n \times n}$ such that $B=QQ^\top$. Hence $\Tr AB = \Tr AQQ^\top = \Tr Q^\top AQ = \sum_{i=1}^n q_i^\top A q_i \geq 0$, where $q_i$ is the $i$-th column of $Q$ and $q_i^\top A q_i \geq 0$ holds because A is positive semidefinite. \\
The proposition now follows immediately.
\end{proof}

\section{Additions to the Diabetes Example}
\label{App:Diabetes}
\begin{figure}
    \centering
    \includegraphics[width=\textwidth]{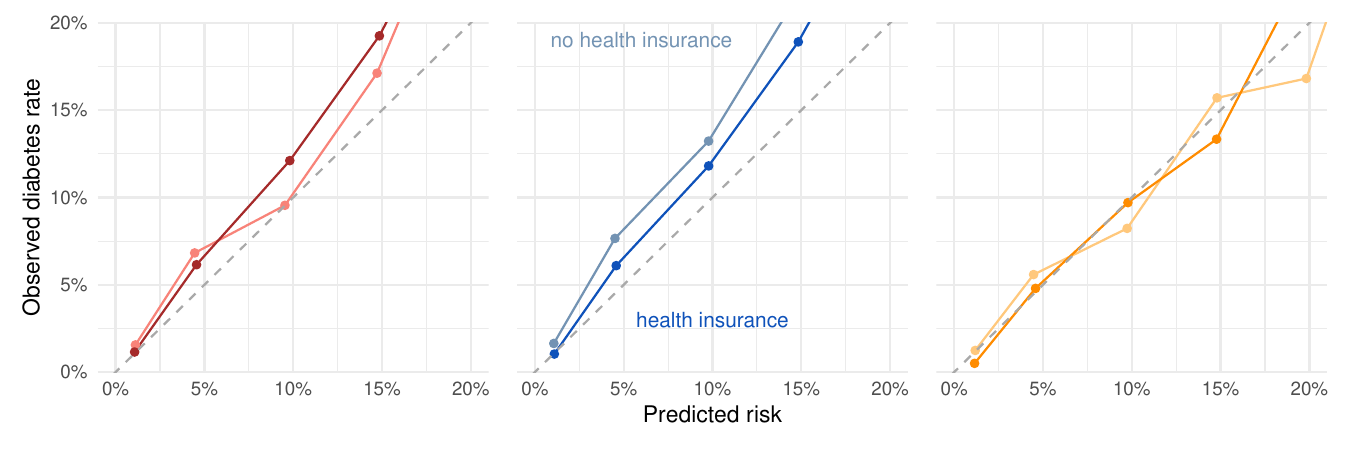}
    \caption{Predicted diabetes risk against diabetes rate against observed diabetes rate estimated for a simple (red), complex (blue), and measurement model (orange) by health insurance status (darker hue: insured, lighter hue: uninsured). For both insured and uninsured people, our measurement model performs better than both regression models closely matching empirical rates of diabetes (dashed gray line).}
    \label{fig:enter-label}
\end{figure}
\subsection{Calculations of the thresholds}
\label{App:threshold}
The thresholds for the measurement model are based purely on information provided in \cite{Fang_Wang_Coresh_Selvin_2022}, especially Tables 1 and 2. We are concerned with the period from 2011 to 2018, the period covered by our NHANES dataset. From \cite{Fang_Wang_Coresh_Selvin_2022}, we know that the rate of total diabetes in 2017--2020 was roughly $14\%$ and that the rate of (persistent) undiagnosed diabetes in people with or without health insurance was roughly $16\%$ and $29\%$, respectively, in 2011--2020. The thresholds $t(\text{health insurance})$ are determined as the shift on the logit scale to match these proportions of undiagnosed patients at the given rate of total diabetes. More concretely, the thresholds can be determined as follows: First, we need to determine a base rate corresponding to $14\%$ total diabetes. Let
\begin{align} \nonumber
    U \sim \text{logistic}(\alpha,1) 
\end{align}
with $\alpha \in \R$, such that, 
\begin{align}\nonumber
    P(U \geq 0) = 14\%.
\end{align}
This holds approximately for $\alpha = -1.8$. Based on this base rate, we can determine the thresholds. Let
\begin{align}\nonumber
    Y \sim \text{logistic}(\alpha+t(\text{insurance}) \bold{1}_{\text{insurance}},1),
\end{align}
such that
\begin{align}
    1-\frac{P(Y\geq 0)}{P(U \geq 0)} = \begin{cases}
        16\% \text{ if insured} \\
        29\% \text{ if uninsured}.
    \end{cases}
\end{align}
In our model, we have coded \textit{insured} as the base level,i.e. $\bold{1}_{\text{insurance}} = 1$ if and only if the patient is uninsured. With this choice, the above holds approximately for $t(uninsured) = -0.38$ and  $t(insured) = -0.21$.
\subsubsection{Further comparisons of prediction quality}
\renewrobustcmd{\bfseries}{\fontseries{b}\selectfont}
\renewrobustcmd{\boldmath}{}
\newrobustcmd{\B}{\bfseries}
\begin{table}
    \centering
    \begin{tabular}{lcccc}
                 &Simple Model & Complex Model & Measurement Model & Oracle Model\\
    \hline
        Log Score & -0.333 & -0.333 & \B-0.324 & -0.324 \\
        Brier Score & -0.206  &  -0.206 & \B-0.202 & -0.202\\
        MSE &  0.014& 0.014 & \B 0.010 & 0.011 \\
        Accuracy & 0.858 & 0.858 &\B 0.862 & 0.861\\
        PPV & 0.574 & 0.573 &\B 0.585 & 0.589\\
        NPV &  0.866 & 0.866 & \B0.875 & 0.872 \\
    \end{tabular}
    \caption{Comparison of the simple, complex, and measurement model across a range of performance metrics for classification. The measurement model outperforms the classical logistics regression models throughout and is similar in performance to an Oracle model trained on the true labels. }
    \label{tab:app_diabetes}
\end{table}
Table \ref{tab:app_diabetes} shows comparisons between the simple, complex, and measurement models in terms of a variety of metrics for classification quality. The measurement model improves on both the simple and complex logistic regression model with a performance that is on par with an oracle logistic regression model trained on the true labels.

Both the log and Brier scores are strictly proper scoring rules. Scoring rules are summary measures for the quality of probabilistic predictions for classification, which take both accuracy and calibration into account. More explicitly, a scoring rule $S(x,Q)$ measures the quality of the distribution $Q$ for predicting a discrete random quantity $X$, when $X=x$ is observed. A scoring rule is called proper, if the expected score $E_{X\sim P}S(X,Q)$ is (strictly) minimized by $Q=P$ \citep{Gneiting_Raftery_2007}. The log score is defined by setting $S(x,Q) = \log Q(x)$. The Brier score is obtained with $S(x,Q) = 2Q(x) - \sum_{j=1}^m Q(j)^2 -1$, where the sum runs over all classes ($m=2$ in our diabetes example).
    
Accuracy is defined as the proportion of patients correctly classified. 
Positive predictive value (PPV) is the probability that a patient classified with diabetes actually has the disease. Similarly, negative predictive value (NPV) is the probability that a patient predicted not to have diabetes is actually free of diabetes. 
\section{Determining parameters that require strong priors}
\label{App:DetParamsStrongPri}
When the relationship between proxy and true outcomes is unknown, Bayesian measurement models are only partially identified \citep{Gustafson_2015}. This necessitates strong priors or treating non-identified parameters as sensitivity parameters in a sensitivity analysis (see Section \ref{Sec:Miss}).
This section briefly outlines potential ways of determining parameters that are not informed by data and hence require strong priors.

One way to single out parameters that require strong priors is transparent parameterization \citep{Gustafson_2009,Richardson_Evans_Robins_2011}, in which the model is reparameterized to separate point-identified parameters and completely non-identified parameters. The latter require strong priors and should be treated as sensitivity parameters in a sensitivity analysis. 

We demonstrate this approach for the leakage model used for the stylized example of criminal behavior and arrests (see Section \ref{Sec:leakageModel}).
We seek to re-parameterize our model (equation \eqref{Eq:leakage}) from $\lambda = \{\alpha,\beta,\gamma,\eta,\sigma_y^2\}$ to $(\phi,\psi)$, such that the distribution of $(y_0,y_1)$ depends only on $\phi$ and not on any lower-dimensional function of it. 
For this model, it is straightforward to marginalize out the true outcomes $u_0$ and $u_1$ to arrive at
\begin{align}
     \begin{bmatrix}
y_0  \\
y_1   
\end{bmatrix} &\sim \text{MVN}\bigg(\begin{bmatrix}
(\alpha+\beta\gamma)  X  \\
(\alpha+\beta\gamma) X   
\end{bmatrix},\begin{bmatrix}
\sigma_y^2 & \sigma_u^2 \eta \\
\sigma_u^2 \eta& \sigma_y^2 
\end{bmatrix}\bigg). 
\end{align}
Clearly, the set of parameters $\phi=\{(\alpha+\beta\gamma),\sigma_y^2,\sigma_u^2\eta\}$ is minimal sufficient for $(y_0,y_1)$. These parameters are point-identified. We collect the remaining parameters $\psi = \{\beta,\gamma\}$ and have, by construction, that $\psi|\phi, y_0, y_1 = \psi|\phi$ does not depend on the data and is thus sensitive to the prior conditional on $\psi|\phi$.

While this approach is compelling, it requires analytical derivations, and not every model is guaranteed to afford such a transparent parameterization. A more general approach is based on prior sensitivity analysis which checks the sensitivity of the posterior to changes in the prior. This is an intuitive notion of identification of parameters in the Bayesian paradigm \citep{Li_Ding_Mealli_2022}. 
While prior sensitivity analysis can be performed naively, there is ongoing research on using it in a computationally efficient manner \citep{Depaoli_Winter_Visser_2020,Gelman_Vehtari_Simpson_Margossian_Carpenter_Yao_Kennedy_Gabry_Bürkner_Modrák_2020}. \citet{Kallioinen_Paananen_Bürkner_Vehtari_2024} present an approach based on power-scaling via importance sampling that is able to identify the set of parameters of our leakage model that require a strong prior. 
\end{document}